\documentstyle[twocolumn,prl,aps]{revtex}
\def\PsfigVersion{1.9}
\ifx\undefined\psfig\else \fi

\let\LaTeXAtSign=\@
\let\@=\relax
\edef\psfigRestoreAt{\catcode`\@=\number\catcode`@\relax}
\catcode`\@=11\relax
\newwrite\@unused
\def\ps@typeout#1{{\let\protect\string\immediate\write\@unused{#1}}}
\ps@typeout{psfig/tex \PsfigVersion}

\def\figurepath{./}

\def\@nnil{\@nil}
\def\@empty{}
\def\@psdonoop#1\@@#2#3{}
\def\@psdo#1:=#2\do#3{\edef\@psdotmp{#2}\ifx\@psdotmp\@empty \else
    \expandafter\@psdoloop#2,\@nil,\@nil\@@#1{#3}\fi}
\def\@psdoloop#1,#2,#3\@@#4#5{\def#4{#1}\ifx #4\@nnil \else
       #5\def#4{#2}\ifx #4\@nnil \else#5\@ipsdoloop #3\@@#4{#5}\fi\fi}
\def\@ipsdoloop#1,#2\@@#3#4{\def#3{#1}\ifx #3\@nnil 
       \let\@nextwhile=\@psdonoop \else
      #4\relax\let\@nextwhile=\@ipsdoloop\fi\@nextwhile#2\@@#3{#4}}
\def\@tpsdo#1:=#2\do#3{\xdef\@psdotmp{#2}\ifx\@psdotmp\@empty \else
    \@tpsdoloop#2\@nil\@nil\@@#1{#3}\fi}
\def\@tpsdoloop#1#2\@@#3#4{\def#3{#1}\ifx #3\@nnil 
       \let\@nextwhile=\@psdonoop \else
      #4\relax\let\@nextwhile=\@tpsdoloop\fi\@nextwhile#2\@@#3{#4}}
\ifx\undefined\fbox
\newdimen\fboxrule
\newdimen\fboxsep
\newdimen\ps@tempdima
\newbox\ps@tempboxa
\long\def\fbox#1{\leavevmode\setbox\ps@tempboxa\hbox{#1}\ps@tempdima\fboxrule
    \advance\ps@tempdima \fboxsep \advance\ps@tempdima \dp\ps@tempboxa
   \hbox{\lower \ps@tempdima\hbox
  {\vbox{\hrule height \fboxrule
          \hbox{\vrule width \fboxrule \hskip\fboxsep
          \vbox{\vskip\fboxsep \box\ps@tempboxa\vskip\fboxsep}\hskip 
                 \fboxsep\vrule width \fboxrule}
                 \hrule height \fboxrule}}}}
\fi
\newread\ps@stream
\newif\ifnot@eof       
\newif\if@noisy        
\newif\if@atend        
\newif\if@psfile       
{\catcode`\%=12\global\gdef\epsf@start{
\def\epsf@PS{PS}
\def\epsf@getbb#1{%
\openin\ps@stream=#1
\ifeof\ps@stream\ps@typeout{Error, File #1 not found}\else
   {\not@eoftrue \chardef\other=12
    \def\do##1{\catcode`##1=\other}\dospecials \catcode`\ =10
    \loop
       \if@psfile
	  \read\ps@stream to \epsf@fileline
       \else{
	  \obeyspaces
          \read\ps@stream to \epsf@tmp\global\let\epsf@fileline\epsf@tmp}
       \fi
       \ifeof\ps@stream\not@eoffalse\else
       \if@psfile\else
       \expandafter\epsf@test\epsf@fileline:. \\%
       \fi
          \expandafter\epsf@aux\epsf@fileline:. \\%
       \fi
   \ifnot@eof\repeat
   }\closein\ps@stream\fi}%
\long\def\epsf@test#1#2#3:#4\\{\def\epsf@testit{#1#2}
			\ifx\epsf@testit\epsf@start\else
\ps@typeout{Warning! File does not start with `\epsf@start'.  It may not be a PostScript file.}
			\fi
			\@psfiletrue} 
{\catcode`\%=12\global\let\epsf@percent=
\long\def\epsf@aux#1#2:#3\\{\ifx#1\epsf@percent
   \def\epsf@testit{#2}\ifx\epsf@testit\epsf@bblit
	\@atendfalse
        \epsf@atend #3 . \\%
	\if@atend	
	   \if@verbose{
		\ps@typeout{psfig: found `(atend)'; continuing search}
	   }\fi
        \else
        \epsf@grab #3 . . . \\%
        \not@eoffalse
        \global\no@bbfalse
        \fi
   \fi\fi}%
\def\epsf@grab #1 #2 #3 #4 #5\\{%
   \global\def\epsf@llx{#1}\ifx\epsf@llx\empty
      \epsf@grab #2 #3 #4 #5 .\\\else
   \global\def\epsf@lly{#2}%
   \global\def\epsf@urx{#3}\global\def\epsf@ury{#4}\fi}%
\def\epsf@atendlit{(atend)} 
\def\epsf@atend #1 #2 #3\\{%
   \def\epsf@tmp{#1}\ifx\epsf@tmp\empty
      \epsf@atend #2 #3 .\\\else
   \ifx\epsf@tmp\epsf@atendlit\@atendtrue\fi\fi}

\chardef\psletter = 11 
\chardef\other = 12

\newif \ifdebug 
\newif\ifc@mpute 
\c@mputetrue 

\let\then = \relax
\def\r@dian{pt }
\let\r@dians = \r@dian
\let\dimensionless@nit = \r@dian
\let\dimensionless@nits = \dimensionless@nit
\def\internal@nit{sp }
\let\internal@nits = \internal@nit
\newif\ifstillc@nverging
\def \Mess@ge #1{\ifdebug \then \message {#1} \fi}

{ 
	\catcode `\@ = \psletter
	\gdef \nodimen {\expandafter \n@dimen \the \dimen}
	\gdef \term #1 #2 #3%
	       {\edef \t@ {\the #1}
		\edef \t@@ {\expandafter \n@dimen \the #2\r@dian}%
		\t@rm {\t@} {\t@@} {#3}%
	       }
	\gdef \t@rm #1 #2 #3%
	       {{%
		\count 0 = 0
		\dimen 0 = 1 \dimensionless@nit
		\dimen 2 = #2\relax
		\Mess@ge {Calculating term #1 of \nodimen 2}%
		\loop
		\ifnum	\count 0 < #1
		\then	\advance \count 0 by 1
			\Mess@ge {Iteration \the \count 0 \space}%
			\Multiply \dimen 0 by {\dimen 2}%
			\Mess@ge {After multiplication, term = \nodimen 0}%
			\Divide \dimen 0 by {\count 0}%
			\Mess@ge {After division, term = \nodimen 0}%
		\repeat
		\Mess@ge {Final value for term #1 of 
				\nodimen 2 \space is \nodimen 0}%
		\xdef \Term {#3 = \nodimen 0 \r@dians}%
		\aftergroup \Term
	       }}
	\catcode `\p = \other
	\catcode `\t = \other
	\gdef \n@dimen #1pt{#1} 
}

\def \Divide #1by #2{\divide #1 by #2} 

\def \Multiply #1by #2
       {{
	\count 0 = #1\relax
	\count 2 = #2\relax
	\count 4 = 65536
	\Mess@ge {Before scaling, count 0 = \the \count 0 \space and
			count 2 = \the \count 2}%
	\ifnum	\count 0 > 32767 
	\then	\divide \count 0 by 4
		\divide \count 4 by 4
	\else	\ifnum	\count 0 < -32767
		\then	\divide \count 0 by 4
			\divide \count 4 by 4
		\else
		\fi
	\fi
	\ifnum	\count 2 > 32767 
	\then	\divide \count 2 by 4
		\divide \count 4 by 4
	\else	\ifnum	\count 2 < -32767
		\then	\divide \count 2 by 4
			\divide \count 4 by 4
		\else
		\fi
	\fi
	\multiply \count 0 by \count 2
	\divide \count 0 by \count 4
	\xdef \product {#1 = \the \count 0 \internal@nits}%
	\aftergroup \product
       }}

\def\r@duce{\ifdim\dimen0 > 90\r@dian \then   
		\multiply\dimen0 by -1
		\advance\dimen0 by 180\r@dian
		\r@duce
	    \else \ifdim\dimen0 < -90\r@dian \then  
		\advance\dimen0 by 360\r@dian
		\r@duce
		\fi
	    \fi}

\def\Sine#1%
       {{%
	\dimen 0 = #1 \r@dian
	\r@duce
	\ifdim\dimen0 = -90\r@dian \then
	   \dimen4 = -1\r@dian
	   \c@mputefalse
	\fi
	\ifdim\dimen0 = 90\r@dian \then
	   \dimen4 = 1\r@dian
	   \c@mputefalse
	\fi
	\ifdim\dimen0 = 0\r@dian \then
	   \dimen4 = 0\r@dian
	   \c@mputefalse
	\fi
	\ifc@mpute \then
		\divide\dimen0 by 180
		\dimen0=3.141592654\dimen0
		\dimen 2 = 3.1415926535897963\r@dian 
		\divide\dimen 2 by 2 
		\Mess@ge {Sin: calculating Sin of \nodimen 0}%
		\count 0 = 1 
		\dimen 2 = 1 \r@dian 
		\dimen 4 = 0 \r@dian 
		\loop
			\ifnum	\dimen 2 = 0 
			\then	\stillc@nvergingfalse 
			\else	\stillc@nvergingtrue
			\fi
			\ifstillc@nverging 
			\then	\term {\count 0} {\dimen 0} {\dimen 2}%
				\advance \count 0 by 2
				\count 2 = \count 0
				\divide \count 2 by 2
				\ifodd	\count 2 
				\then	\advance \dimen 4 by \dimen 2
				\else	\advance \dimen 4 by -\dimen 2
				\fi
		\repeat
	\fi		
			\xdef \sine {\nodimen 4}%
       }}

\def\Cosine#1{\ifx\sine\UnDefined\edef\Savesine{\relax}\else
		             \edef\Savesine{\sine}\fi
	{\dimen0=#1\r@dian\advance\dimen0 by 90\r@dian
	 \Sine{\nodimen 0}
	 \xdef\cosine{\sine}
	 \xdef\sine{\Savesine}}}	      

\def\psdraft{
	\def\@psdraft{0}
}
\def\psfull{
	\def\@psdraft{100}
}

\psfull

\newif\if@scalefirst
\def\psscalefirst{\@scalefirsttrue}
\def\psrotatefirst{\@scalefirstfalse}
\psrotatefirst

\newif\if@draftbox
\def\psnodraftbox{
	\@draftboxfalse
}
\def\psdraftbox{
	\@draftboxtrue
}
\@draftboxtrue

\newif\if@prologfile
\newif\if@postlogfile
\def\pssilent{
	\@noisyfalse
}
\def\psnoisy{
	\@noisytrue
}
\psnoisy
\newif\if@bbllx
\newif\if@bblly
\newif\if@bburx
\newif\if@bbury
\newif\if@height
\newif\if@width
\newif\if@rheight
\newif\if@rwidth
\newif\if@angle
\newif\if@clip
\newif\if@verbose
\def\@p@@sclip#1{\@cliptrue}

\newif\if@decmpr

\def\@p@@sfigure#1{\def\@p@sfile{null}\def\@p@sbbfile{null}
	        \openin1=#1
		\ifeof1\closein1
	        	\openin1=\figurepath#1
			\ifeof1\closein1
			        \openin1=#1
				\ifeof1\closein1%
				       \openin1=\figurepath#1
					\ifeof1
					   \ps@typeout{Error, File #1 not found}
						\if@bbllx\if@bblly
				   		\if@bburx\if@bbury
			      				\def\@p@sfile{#1}%
			      				\def\@p@sbbfile{#1}%
							\@decmprfalse
				  	   	\fi\fi\fi\fi
					\else\closein1
				    		\def\@p@sfile{\figurepath#1}%
				    		\def\@p@sbbfile{\figurepath#1}%
						\@decmprfalse
	                       		\fi%
			 	\else\closein1%
					\def\@p@sfile{#1}
					\def\@p@sbbfile{#1}
					\@decmprfalse
			 	\fi
			\else
				\def\@p@sfile{\figurepath#1}
				\def\@p@sbbfile{\figurepath#1}
				\@decmprtrue
			\fi
		\else
			\def\@p@sfile{#1}
			\def\@p@sbbfile{#1}
			\@decmprtrue
		\fi}

\def\@p@@sfile#1{\@p@@sfigure{#1}}

\def\@p@@sbbllx#1{
		\@bbllxtrue
		\dimen100=#1
		\edef\@p@sbbllx{\number\dimen100}
}
\def\@p@@sbblly#1{
		\@bbllytrue
		\dimen100=#1
		\edef\@p@sbblly{\number\dimen100}
}
\def\@p@@sbburx#1{
		\@bburxtrue
		\dimen100=#1
		\edef\@p@sbburx{\number\dimen100}
}
\def\@p@@sbbury#1{
		\@bburytrue
		\dimen100=#1
		\edef\@p@sbbury{\number\dimen100}
}
\def\@p@@sheight#1{
		\@heighttrue
		\dimen100=#1
   		\edef\@p@sheight{\number\dimen100}
}
\def\@p@@swidth#1{
		\@widthtrue
		\dimen100=#1
		\edef\@p@swidth{\number\dimen100}
}
\def\@p@@srheight#1{
		\@rheighttrue
		\dimen100=#1
		\edef\@p@srheight{\number\dimen100}
}
\def\@p@@srwidth#1{
		\@rwidthtrue
		\dimen100=#1
		\edef\@p@srwidth{\number\dimen100}
}
\def\@p@@sangle#1{
		\@angletrue
		\edef\@p@sangle{#1} 
}
\def\@p@@ssilent#1{ 
		\@verbosefalse
}
\def\@p@@sprolog#1{\@prologfiletrue\def\@prologfileval{#1}}
\def\@p@@spostlog#1{\@postlogfiletrue\def\@postlogfileval{#1}}
\def\@cs@name#1{\csname #1\endcsname}
\def\@setparms#1=#2,{\@cs@name{@p@@s#1}{#2}}
\def\ps@init@parms{
		\@bbllxfalse \@bbllyfalse
		\@bburxfalse \@bburyfalse
		\@heightfalse \@widthfalse
		\@rheightfalse \@rwidthfalse
		\def\@p@sbbllx{}\def\@p@sbblly{}
		\def\@p@sbburx{}\def\@p@sbbury{}
		\def\@p@sheight{}\def\@p@swidth{}
		\def\@p@srheight{}\def\@p@srwidth{}
		\def\@p@sangle{0}
		\def\@p@sfile{} \def\@p@sbbfile{}
		\def\@p@scost{10}
		\def\@sc{}
		\@prologfilefalse
		\@postlogfilefalse
		\@clipfalse
		\if@noisy
			\@verbosetrue
		\else
			\@verbosefalse
		\fi
}
\def\parse@ps@parms#1{
	 	\@psdo\@psfiga:=#1\do
		   {\expandafter\@setparms\@psfiga,}}
\newif\ifno@bb
\def\bb@missing{
	\if@verbose{
		\ps@typeout{psfig: searching \@p@sbbfile \space  for bounding box}
	}\fi
	\no@bbtrue
	\epsf@getbb{\@p@sbbfile}
        \ifno@bb \else \bb@cull\epsf@llx\epsf@lly\epsf@urx\epsf@ury\fi
}	
\def\bb@cull#1#2#3#4{
	\dimen100=#1 bp\edef\@p@sbbllx{\number\dimen100}
	\dimen100=#2 bp\edef\@p@sbblly{\number\dimen100}
	\dimen100=#3 bp\edef\@p@sbburx{\number\dimen100}
	\dimen100=#4 bp\edef\@p@sbbury{\number\dimen100}
	\no@bbfalse
}
\newdimen\p@intvaluex
\newdimen\p@intvaluey
\def\rotate@#1#2{{\dimen0=#1 sp\dimen1=#2 sp
		  \global\p@intvaluex=\cosine\dimen0
		  \dimen3=\sine\dimen1
		  \global\advance\p@intvaluex by -\dimen3
		  \global\p@intvaluey=\sine\dimen0
		  \dimen3=\cosine\dimen1
		  \global\advance\p@intvaluey by \dimen3
		  }}
\def\compute@bb{
		\no@bbfalse
		\if@bbllx \else \no@bbtrue \fi
		\if@bblly \else \no@bbtrue \fi
		\if@bburx \else \no@bbtrue \fi
		\if@bbury \else \no@bbtrue \fi
		\ifno@bb \bb@missing \fi
		\ifno@bb \ps@typeout{FATAL ERROR: no bb supplied or found}
			\no-bb-error
		\fi
		\count203=\@p@sbburx
		\count204=\@p@sbbury
		\advance\count203 by -\@p@sbbllx
		\advance\count204 by -\@p@sbblly
		\edef\ps@bbw{\number\count203}
		\edef\ps@bbh{\number\count204}
		\if@angle 
			\Sine{\@p@sangle}\Cosine{\@p@sangle}
	        	{\dimen100=\maxdimen\xdef\r@p@sbbllx{\number\dimen100}
					    \xdef\r@p@sbblly{\number\dimen100}
			                    \xdef\r@p@sbburx{-\number\dimen100}
					    \xdef\r@p@sbbury{-\number\dimen100}}
                        \def\minmaxtest{
			   \ifnum\number\p@intvaluex<\r@p@sbbllx
			      \xdef\r@p@sbbllx{\number\p@intvaluex}\fi
			   \ifnum\number\p@intvaluex>\r@p@sbburx
			      \xdef\r@p@sbburx{\number\p@intvaluex}\fi
			   \ifnum\number\p@intvaluey<\r@p@sbblly
			      \xdef\r@p@sbblly{\number\p@intvaluey}\fi
			   \ifnum\number\p@intvaluey>\r@p@sbbury
			      \xdef\r@p@sbbury{\number\p@intvaluey}\fi
			   }
			\rotate@{\@p@sbbllx}{\@p@sbblly}
			\minmaxtest
			\rotate@{\@p@sbbllx}{\@p@sbbury}
			\minmaxtest
			\rotate@{\@p@sbburx}{\@p@sbblly}
			\minmaxtest
			\rotate@{\@p@sbburx}{\@p@sbbury}
			\minmaxtest
			\edef\@p@sbbllx{\r@p@sbbllx}\edef\@p@sbblly{\r@p@sbblly}
			\edef\@p@sbburx{\r@p@sbburx}\edef\@p@sbbury{\r@p@sbbury}
		\fi
		\count203=\@p@sbburx
		\count204=\@p@sbbury
		\advance\count203 by -\@p@sbbllx
		\advance\count204 by -\@p@sbblly
		\edef\@bbw{\number\count203}
		\edef\@bbh{\number\count204}
}
\def\in@hundreds#1#2#3{\count240=#2 \count241=#3
		     \count100=\count240	
		     \divide\count100 by \count241
		     \count101=\count100
		     \multiply\count101 by \count241
		     \advance\count240 by -\count101
		     \multiply\count240 by 10
		     \count101=\count240	
		     \divide\count101 by \count241
		     \count102=\count101
		     \multiply\count102 by \count241
		     \advance\count240 by -\count102
		     \multiply\count240 by 10
		     \count102=\count240	
		     \divide\count102 by \count241
		     \count200=#1\count205=0
		     \count201=\count200
			\multiply\count201 by \count100
		 	\advance\count205 by \count201
		     \count201=\count200
			\divide\count201 by 10
			\multiply\count201 by \count101
			\advance\count205 by \count201
		     \count201=\count200
			\divide\count201 by 100
			\multiply\count201 by \count102
			\advance\count205 by \count201
		     \edef\@result{\number\count205}
}
\def\compute@wfromh{
		\in@hundreds{\@p@sheight}{\@bbw}{\@bbh}
		\edef\@p@swidth{\@result}
}
\def\compute@hfromw{
	        \in@hundreds{\@p@swidth}{\@bbh}{\@bbw}
		\edef\@p@sheight{\@result}
}
\def\compute@handw{
		\if@height 
			\if@width
			\else
				\compute@wfromh
			\fi
		\else 
			\if@width
				\compute@hfromw
			\else
				\edef\@p@sheight{\@bbh}
				\edef\@p@swidth{\@bbw}
			\fi
		\fi
}
\def\compute@resv{
		\if@rheight \else \edef\@p@srheight{\@p@sheight} \fi
		\if@rwidth \else \edef\@p@srwidth{\@p@swidth} \fi
}
\def\compute@sizes{
	\compute@bb
	\if@scalefirst\if@angle
	\if@width
	   \in@hundreds{\@p@swidth}{\@bbw}{\ps@bbw}
	   \edef\@p@swidth{\@result}
	\fi
	\if@height
	   \in@hundreds{\@p@sheight}{\@bbh}{\ps@bbh}
	   \edef\@p@sheight{\@result}
	\fi
	\fi\fi
	\compute@handw
	\compute@resv}

\def\psfig#1{\vbox {
	\ps@init@parms
	\parse@ps@parms{#1}
	\compute@sizes
	\ifnum\@p@scost<\@psdraft{
		\special{ps::[begin] 	\@p@swidth \space \@p@sheight \space
				\@p@sbbllx \space \@p@sbblly \space
				\@p@sbburx \space \@p@sbbury \space
				startTexFig \space }
		\if@angle
			\special {ps:: \@p@sangle \space rotate \space} 
		\fi
		\if@clip{
			\if@verbose{
				\ps@typeout{(clip)}
			}\fi
			\special{ps:: doclip \space }
		}\fi
		\if@prologfile
		    \special{ps: plotfile \@prologfileval \space } \fi
		\if@decmpr{
			\if@verbose{
				\ps@typeout{psfig: including \@p@sfile \space }
			}\fi
			\special{ps: plotfile \@p@sfile \space }
		}\else{
			\if@verbose{
				\ps@typeout{psfig: including \@p@sfile \space }
			}\fi
			\special{ps: plotfile \@p@sfile \space }
		}\fi
		\if@postlogfile
		    \special{ps: plotfile \@postlogfileval \space } \fi
		\special{ps::[end] endTexFig \space }
		\vbox to \@p@srheight sp{
			\hbox to \@p@srwidth sp{
				\hss
			}
		\vss
		}
	}\else{
		\if@draftbox{		
			\hbox{\frame{\vbox to \@p@srheight sp{
			\vss
			\hbox to \@p@srwidth sp{ \hss \@p@sfile \hss }
			\vss
			}}}
		}\else{
			\vbox to \@p@srheight sp{
			\vss
			\hbox to \@p@srwidth sp{\hss}
			\vss
			}
		}\fi

	}\fi
}}
\psfigRestoreAt
\let\@=\LaTeXAtSign

\tightenlines
\begin{document}

\twocolumn[
\hsize\textwidth\columnwidth\hsize\csname@twocolumnfalse\endcsname

\draft

\title{Selection Rules for Resonant Inelastic X-Ray Scattering from Tetragonal Copper-Oxides}

\author{P.~Abbamonte$^{1,2,*}$, C.~A.~Burns$^3$, E.~D.~Isaacs$^2$,
P.~M.~Platzman$^2$, L.~L.~Miller$^4$, and M.~V.~Klein$^1$\\}
 
\address{
$^1$Department of Physics, University of Illinois, 1110 W. Green St., Urbana, IL, 61801\\
$^2$Bell Laboratories, Lucent Technologies, 600 Mountain Av., Murray Hill, NJ, 07974\\
$^3$Department of Physics, Western Michigan University, Kalamazoo, MI, 49008\\
$^4$Ames Laboratory, Ames, IA, 50011\\
}
 
\date{\today}
\maketitle
 
\begin{abstract}
We demonstrate the utility of point group representation theory 
for symmetry analysis in
resonant inelastic x-ray scattering.  From its 
polarization-dependence, we show that a
5 eV inelastic feature in Sr$_2$CuO$_2$Cl$_2$ has pure $B_{1g}$ 
symmetry and assign it to
a transition in the cell-perturbation calculations of 
S\'{\i}mon {\it et. al.} {[Phys. Rev. B., {\bf 54}, R3780 (1996)]}.  
We discuss how Raman selection rules 
are broken at nonzero momentum transfer and how this can also act as a
probe of wave function symmetry.
\end{abstract}
 
]
\narrowtext
The two advantages of resonant as compared to 
nonresonant inelastic x-ray scattering
are that it can be applied to high-density materials 
(where nonresonant techniques
have problems with absorption) and its 
sensitivty to wave function {\it symmetry}.
The former is well-documented but the latter has 
only recently been applied to solids by
Duda\cite{nordgrenJJAP} and Kuiper\cite{nordgrenPRL}.

In this article we demonstrate the utility of point group representation
theory for symmetry analysis in resonant inelastic x-ray scattering (RIXS).  
We focus on 
inelastic features which are brought
about by the coulomb interaction between core and valence electrons.  
In different contexts such features have been called ``forbidden 
excitonic" transitions\cite{vanveenendaal}, ``indirect"
transitions\cite{gelmukhanov}, 
``Coster-Kronig" features\cite{ederer}, ``Auger resonant 
Raman" features\cite{karis},
and ``shakeup" features\cite{phil,me}.  One expects these transitions 
to dominate K-edge RIXS spectra
from $d$-electron systems with inversion symmetry 
(i.e. many transition metal oxides)
owing to the absence of dipole-allowed transitions 
at the Brillouin zone center.
We take the ``shakeup" approach of Refs.\cite{phil,me} and tabulate the Raman-active
symmetries in all independent experimental geometries.  Comparing our polarization-dependent
measurements of the 5 eV inelastic feature in Sr$_2$CuO$_2$Cl$_2$ (SCOC) 
(observed also in Nd$_2$CuO$_4$\cite{johnhill} and La$_2$CuO$_4$\cite{me})
with the cell-perturbation
calculations of S\'{\i}mon\cite{simon} we identify this feature 
to be a localized transition of $B_{1g}$ (or $d_{x^2-y^2}$) symmetry.  
We conclude by discussing
how selection rules are broken at nonzero momentum transfer and how this
can act as an additional probe of wave function symmetry.

Experiments were carried out at the 3ID (SRI-CAT) beam line at the 
Advanced Photon Source using a 6-circle diffractometer with an
analyzer stage.  Energy analysis of the scattered 
light was done with a spherical 
Ge(733) analyzer working near backscattering.  The overall energy resolution
was 0.9 eV, and with $6\times 10^{12}$ phonons/sec on the sample typical inelastic
count rates were 10 Hz.  The scattering angle in the experiment was fixed at 
$2\theta=16^o$, allowing detection of both parallel and crossed polarization.  
Two detectors were employed so fluorescence yield data could be 
taken simultaneously with RIXS measurements.

The crystals were grown as described previously\cite{lance}, 
cut to (100) and (001) surfaces, and polished with a 1 $\mu$m 
AlO film.  This allowed variation of the incident polarization, 
$\hat{\epsilon}_i$, and the momentum transfer, {\bf q}, independently with 
respect to the crystal axes.  The surface quality was verified with Lau\'e photos 
and an FTIR reflectometer.

Figure 1 shows spectra taken with 
$\hat{\epsilon}_i||{\bf \hat{x}}$
and ${\bf q}=(1.27\AA^{-1}){\bf \hat{z}}$.
The incident energy, $\omega_i$, was varied from 8988 eV 
to 9006 eV to generate a resonance
profile.  A feature at 5 eV energy loss
is visible and resonates in a complex fashion.  It sits on
a flat, incoherent, multi-electron/phonon continuum which 
does not change with $\omega_i$.
This continuum is analogous to the selection rule-violating features 
seen in RIXS from molecules\cite{privalov,skytt,gelmukhanovCO2}
which occur because of interference of vibrational modes.  It could
in principle be disposed of by detuning\cite{privalov,skytt,gelmukhanovCO2},
however it is featureless and so
presents no problem for 
data interpretation\cite{ashcroft}.

The coherent feature at 5 eV energy loss has been shown to be an
``indirect" transition\cite{johnhill,me}, and is what we wish to 
symmetry-analyze. We use the ``shakeup" approach of Refs. \cite{phil,me} to
describe the coupling of x-rays to this excitation.  This approach uses
the weakness of the coulomb interaction (the expansion parameter
being $\alpha$=1/137) to describe
the resonance process analytically in perturbation theory, and works as long
as the coulomb interaction brings about no topological change in the $4p$
band states (this is the case if the $4p$ bandwidth is large, as it is
in this case).  We neglect here the detailed energy structure 
of intermediate states
since we plan to make only general symmetry arguments.

According to Ref.\cite{me} the spectra in Figure 1 can be described 
by the expression

\begin{equation}
w_{i\rightarrow f}=\frac{S_K({\bf q},\omega;\hat{\epsilon}_i,\hat{\epsilon}_s)}
{[(\omega_i-E_K)^2+{\gamma_K}^2][(\omega_s-E_K)^2+{\gamma_K}^2],}
\end{equation}

\noindent
where the function $S_K$ is independent of $\omega_i$ 
and the spectral changes in 
Figure 1 are accounted for entirely
by the energy denominators 
[the doublet structure in the open circles
comes from the double denominator\cite{thesis}].
$S_K$ contains information about Raman 
selection rules and has the form\cite{me}

\begin{equation}
S_K({\bf q},\omega;\hat{\epsilon}_i,\hat{\epsilon}_s)=\frac{2\pi}{\hbar}
\sum_f \left| \sum_{\bar{1s}4p} M_{em}M_{coul}M_{abs} \right|^2.
\end{equation}

\noindent
To derive selection rules we discuss the role each of these three
matrix elements plays in the scattering process.  For simplicity we begin in
the optical limit ({\bf q}=0) and later discuss how selection rules
are broken as {\bf q} is increased.

The action of $M_{abs}$ is to annhilate the incident photon 
and create a virtual $\bar{1s}4p$ pair.  
In the dipole approximation
the symmetry axis of the $4p$ is parallel to the incident polarization,
$\hat{\epsilon}_i$ (provided $\hat{\epsilon}_i$ is along a principal axis).  
$M_{em}$ annhilates the $\bar{1s}4p$
state and creates the scattered photon with polarization $\hat{\epsilon}_s$
again parallel to the $4p$.  

The relationship between
$\hat{\epsilon}_i$ and $\hat{\epsilon}_s$ depends on the
action of $M_{coul}$, which determines how the Cu$\bar{1s}4p$ state couples
to the valence electron system.  $M_{coul}$ has the explicit form

\begin{eqnarray}
M_{coul}=&&-e^2\int d{\bf x} \, d{\bf x'} \, 
\frac {\psi_{\bar{1s'}}^*({\bf x'}) \, \psi_{\bar{1s}}({\bf x'})\,
{\psi_f}^*({\bf x}) \, \psi_i({\bf x})}
{\left| {\bf x}-{\bf x'} \right|}\\
&&+e^2\int d{\bf x} \, d{\bf x'} \, \frac{\psi_{4p'}^*({\bf x'}) \, \psi_{4p}({\bf x'})
{\psi_f}^*({\bf x}) \, \psi_i({\bf x})}
{\left| {\bf x}-{\bf x'} \right|}
\end{eqnarray}

\noindent
plus exchange terms.  These integrals describe the coulomb interaction between 
core and valence electrons 
and have the familiar form of nonrelativistic electron-electron 
scattering in the Born approximation.  Raman selection rules
are determined by whether these integrals vanish (or not) by symmetry.

It is straight-forward to show\cite{thesis} that the 
point group symmetries contained in 
the integrands (3-4) are given by the Kr\"onecker product

\begin{equation}
\Gamma(\psi_{4p'}^*\psi_{4p}-\psi_{\bar{1s'}}^*\psi_{\bar{1s}})\otimes
\Gamma(\psi_f^*\psi_i),
\end{equation}

\noindent
where $\Gamma(f)$ represents the symmetry of the function $f$.
According to the fundamental matrix element selection rule theorem\cite{tinkham}
$M_{coul}$ vanishes unless this quantity 
contains the identical representation, $A_{1g}$.
Or equivalently, $M_{coul}$ is nonzero if and
only if the quantities $\Gamma(\psi_f^*\psi_i)$ 
(the overall symmetry of the valence excitation) and 
$\Gamma(\psi_{4p'}^*\psi_{4p}-\psi_{\bar{1s'}}^*\psi_{\bar{1s}})$
(determined entirely by the experimental geometry) share at least one
common symmetry.  In other words - {\it and this is the key point} 
- the Raman active symmetries
for a given experimental geometry are the symmetries contained in the quantity 
$\Gamma(\psi_{4p'}^*\psi_{4p}-\psi_{\bar{1s'}}^*\psi_{\bar{1s}})$, which depends
only on the experimental geometry.

In the optical limit ({\bf q}=0) $\Gamma(\psi_{1s})=a_{1g}$,
since a Cu$1s$ orbital is invariant under all point group operations 
(Cu sits at an inversion center in SCOC).
So the quantity (5) reduces to 
$\Gamma(\psi_{4p'}^*)\otimes\Gamma(\psi_{4p})
+A_{1g}$.  We conclude that 
transitions of $A_{1g}$ symmetry will be Raman-active in all
experimental geometries.  Furthermore,
differences in selection rules between
different experimental geometries are entirely determined by the $4p$.

We now derive the complete
selection rules for one experimental geometry and then
state the results for the others.  Consider the case 
$\hat{\epsilon}_i || {\bf\hat{x}}$ and 
$\hat{\epsilon}_s || {\bf\hat{y}}$.  
In this case the $4p$ is initially oriented along ${\bf\hat{x}}$, 
but by action of $M_{coul}$ it gets rotated
along ${\bf\hat{y}}$.  The symmetries contained in 
$M_{coul}$ in this case are
$\Gamma(4p_y)\otimes\Gamma(4p_x)+A_{1g}$.  $4p_x$ and $4p_y$ have purely
$e_{ux}$ and $e_{uy}$ symmetry, respectively, 
and $e_{ux}\otimes e_{uy}=B_{2g}$.  So the 
Raman active symmetries in this geometry are $A_{1g}+B_{2g}$.
Using this procedure one can generate Raman selection rules
for all independent experimental geometries.  The results are
shown in Table 1.

Returning to the issue of the symmetry of 
the 5 eV inelastic feature, in Fig. 1 the incident 
polarization was $\hat{\bf x}$ and the scattered light was unpolarized, 
so the symmetry reflected in Fig. 1 is 
$A_{1g}+B_{1g}+B_{2g}+E_{gx}$ (see Table 1).

In Figure 2 we show spectra again with $\hat{\epsilon}_i||{\bf\hat{x}}$, 
but with ${\bf q}$ now parallel to ${\bf\hat{y}}$.  
In the {\bf q}=0 limit 
this geometry contains the same Raman-active symmetries
as Figure 1, and in fact the spectra are not observably changed.  
We therefore tentatively assume that we can apply the 
{\bf q}=0 selection rules of
Table 1 to the present measurements, despite our sizeable {\bf q}.  
We will discuss the justification for this assumption in a moment.

In Figure 3 we show measurements with $\hat{\epsilon}_i||{\bf\hat{z}}$ and 
${\bf q}||{\bf\hat{y}}$.  
In this geometry the Raman-active symmetries are
$A_{1g}+E_{gx}$, and here the peak {\it vanishes}.  The fluorescence yield 
(inset, taken {\it in-situ} with a second detector) 
shows a pronounced edge feature, the elastic scattering still
appears in the spectra at the correct energies, and 
the selection rule-violating, incoherent continuum still appears in the
energy-loss part of the spectrum.  However the 5 eV peak is gone.
We conclude that the
5 eV peak is symmetry-forbidden in this geometry, and 
by process of elimination that
it has either $B_{1g}$ or $B_{2g}$ symmetry. 

Armed with some knowledge of its energy and 
symmetry we are prepared to
make a peak assignment.  We appeal to the cell-perturbation calculations
of S\'{\i}mon {\it et. al.}\cite{simon}, 
which is a treatment of the electronic structure of the
undoped CuO$_2$ plane which emphasizes the symmetries of the
states.  We reproduce the central result in Figure 
4.  The ground state has pure $b_{1g}$ symmetry, and starting at
4.3 eV above $E_f$ there is an unbound band of
$a_{1g}$ symmetry.  A transition from the ground state to this band 
has the symmetry $b_{1g}\otimes a_{1g}=B_{1g}$, which with its energy 
is consistent with the experiment.  There are no transitions of $B_{2g}$
symmetry in the vicinity so we conclude that this is our 
peak.

We finish our discussion by describing how a nonzero {\bf q} should violate
the selection rules of Table 1.  For the simplistic case where 
$\psi_{\bar{1s}}$ and $\psi_{4p}$ are tight-binding states
the expression (5) generalizes to

\begin{equation}
\left [ \Gamma(\psi_{4p'}^*\psi_{4p}-\psi_{\bar{1s'}}^*\psi_{\bar{1s}})
\otimes e^{i{\bf q}\cdot{\bf r}} \right ]
\otimes \Gamma(\psi_f^*\psi_i) 
\end{equation}

\noindent
so the symmetries expressed in a given experimental geometry are
those contained in the quantity 
$\Gamma(\psi_{4p'}^*\psi_{4p}-\psi_{\bar{1s'}}^*\psi_{\bar{1s}})\otimes
e^{i{\bf q}\cdot{\bf r}}$.  Additional symmetries are made Raman-active
by the symmetry-reducing effect of the exponential.  For example, in the
case $\hat{\epsilon}_i || \hat{\epsilon}_s || {\bf\hat{z}}$ in the optical
limit the only Raman-active symmetry is $A_{1g}$.  However with a nonzero 
{\bf q}$||{\bf\hat{x}}$ the symmetry is lowered to 
$A_{1g}+B_{1g}+E_{ux}$\cite{obligatoryCarraReference}.
So in general {\bf q} breaks the Raman selection rules - but not 
completely - and in principle can act as an additional probe of wave
function symmetry.

A few words are in order about why we do not observe this expected 
breakdown of selection rules in our measurements, which should show up
perhaps as a difference between Figs. 1 and 2 or more likely as a weak 5 eV intensity 
in Fig. 3.  A fundamental shortcoming of group theory is that, while it
will tell you if a given integral is nonzero, it tells you nothing
about its {\it size}.  RIXS data with improved statistics 
may yet observe such a breakdown.

Two improvements on our study would be, first, to employ a 90$^o$
scattering geometry as in the case of 
Refs.\cite{nordgrenJJAP,nordgrenPRL} to distinguish between the scattered
polarization states. 
This would allow for complete, model-independent determination
of the symmetry of the excitation (provided the effects of {\bf q}, which
becomes sizeable at high scattering angles, remain unimportant).
A second improvement would be to
employ {\it space} group representation theory, since SCOC contains
glide planes whose symmetry is not accounted for in a simple point group
treatment.

We thank A. Bock, M. R\"ubhausen, and D. van der Marel for useful discussions,
and especially G. A. Sawatzky for many heated debates.
This work was supported by the NSF under grant DMR-9705131 and
by the DOE under contract no. W-31-109-ENG-38.

\begin{table}
\caption{
Raman active symmetries for ``indirect" transitions in 
all independent experimental geometries in
the optical limit, {\bf q}=0.  
Symmetries are given both in the Sch\"onflies and spherical harmonic 
notation.
}  
\begin{tabular}{c|cc}
Polarization&Raman Active Symmetries&\\
(incident/scattered)&Sch\"onflies&$Y_{lm}$\\ \hline
${\bf \hat{x}/\hat{x}}$ or ${\bf \hat{y}/\hat{y}}$&$A_{1g}+B_{1g}$&$s+d_{x^2-y^2}$\\
${\bf \hat{z}/\hat{z}}$&$A_{1g}$&$s$\\
${\bf \hat{x}/\hat{y}}$ or ${\bf \hat{y}/\hat{x}}$&$A_{1g}+B_{2g}$&$s+d_{xy}$\\
${\bf \hat{x}/\hat{z}}$ or ${\bf \hat{z}/\hat{x}}$&$A_{1g}+E_{gx}$&$s+d_{xz}$\\
${\bf \hat{y}/\hat{z}}$ or ${\bf \hat{z}/\hat{y}}$&$A_{1g}+E_{gy}$&$s+d_{yz}$\\
\end{tabular}
\end{table}

\pagebreak
  
\begin{figure} 
\begin{center}
  \mbox{\psfig{figure=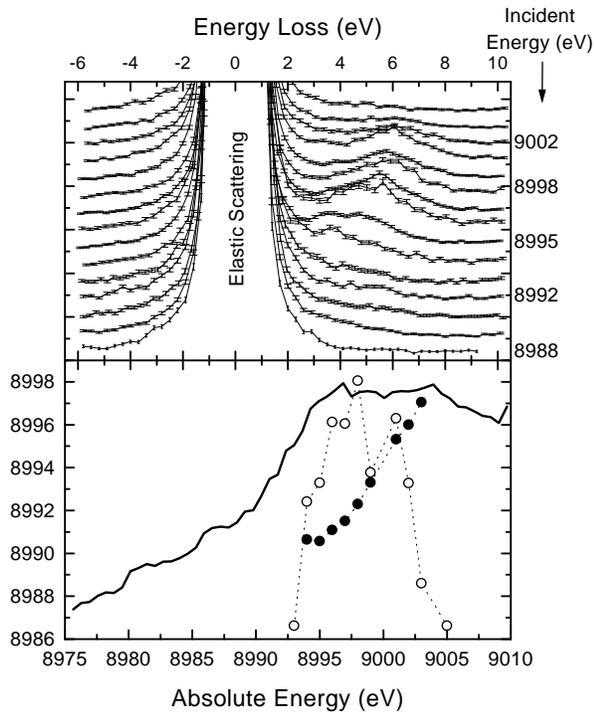,width=3.2in,silent=}}
   \caption{
RIXS spectra from SCOC for different incident energies, $\omega_i$, with 
$\hat{\epsilon}_i||(100)$ and ${\bf q}||(001)$.  The upper panel shows
the individual scans (offset for clarity).  In the lower panel the dark line is the
fluorescence yield, showing the location of the edge.  The open and filled circles are the
inelastic peak height and its energy in absolute units, respectively, plotted against $\omega_i$.
In the {\bf q}=0 limit the Raman-active symmetries in this geometry are $A_{1g}+B_{1g}+B_{2g}+E_{gx}$.
}
\end{center}
\end{figure}

\pagebreak
.
\pagebreak

\begin{figure} 
\begin{center}
  \mbox{\psfig{figure=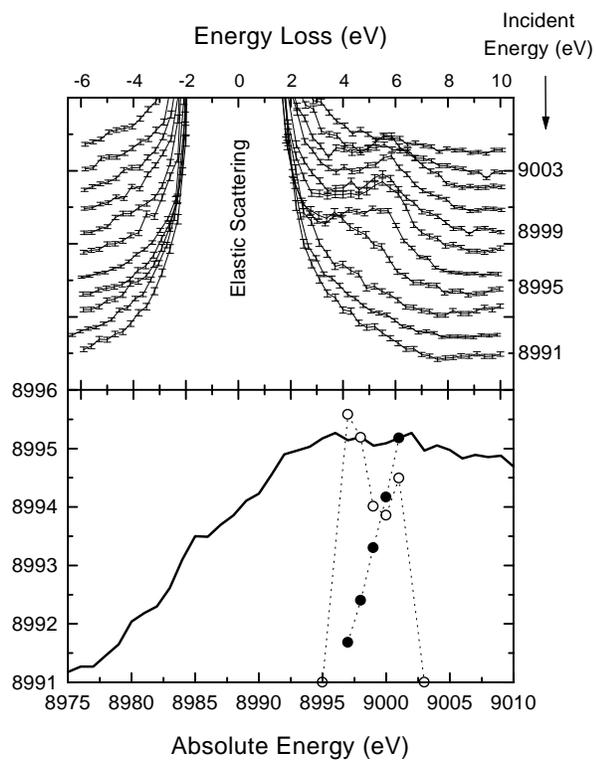,width=3.2in,silent=}}
   \caption{
The same measurement as shown in Fig. 1, however with {\bf q} now directed 
along $\bf{\hat{x}}$.  In the optical limit this geometry contains the 
same symmetries as Fig. 1.
}
\end{center}
\end{figure}

\pagebreak
.
\pagebreak

\begin{figure} 
\begin{center}
  \mbox{\psfig{figure=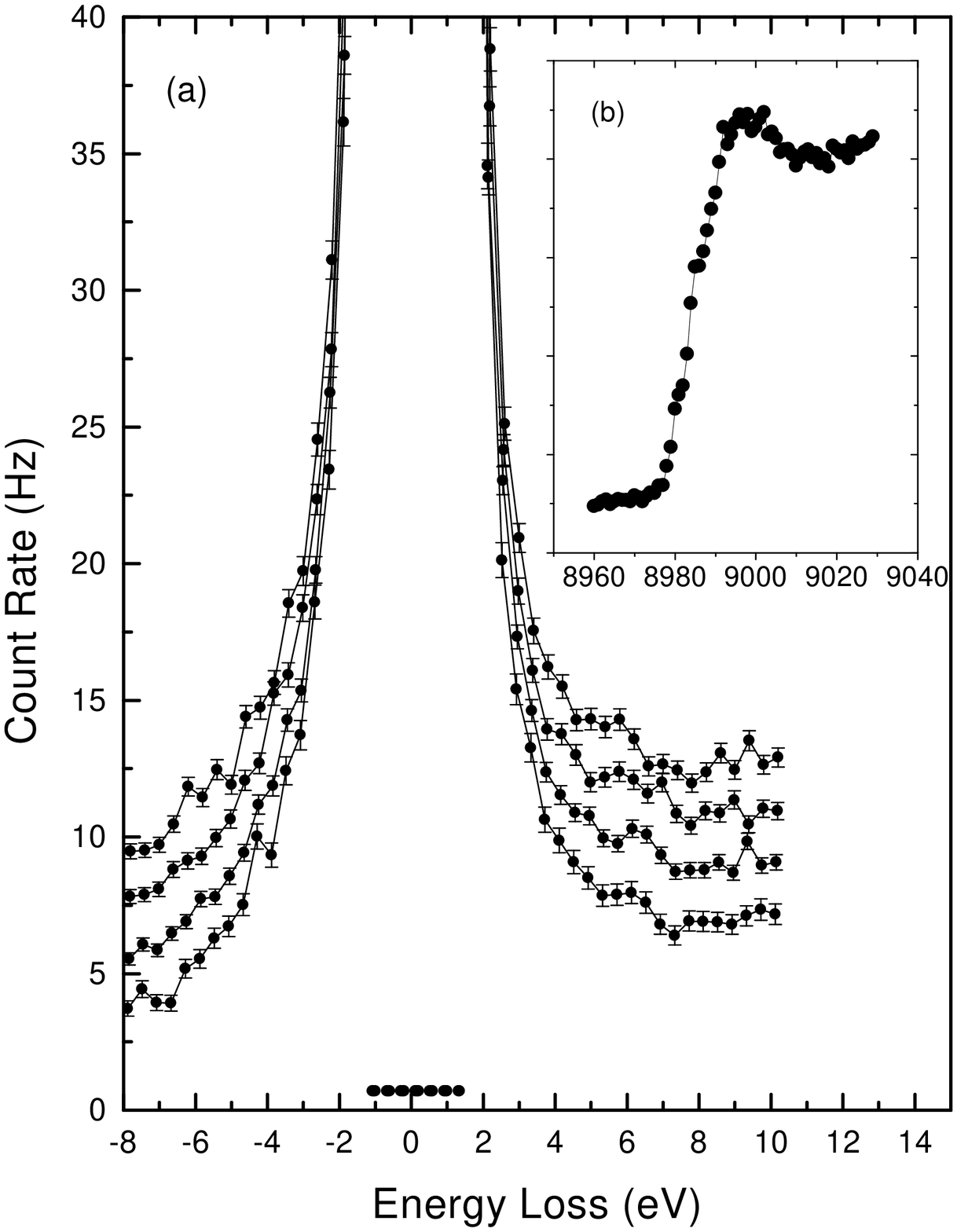,width=3.2 in,silent=}}
   \caption{
The same measurements as in Figs. 1 and 2, however with 
$\hat{\epsilon_i} || {\bf \hat{z}}$.  The fluorescence yield spectrum
taken {\it in situ} is inset.
The Raman-active symmetries
in this geometry are $A_{1g}+E_{gx}$. The 5 eV peak, which
likely has $B_{1g}$ symmetry, is Raman-forbidden.
}
\end{center}
\end{figure}

\pagebreak
.
\pagebreak

\begin{figure} 
\begin{center}
  \mbox{\psfig{figure=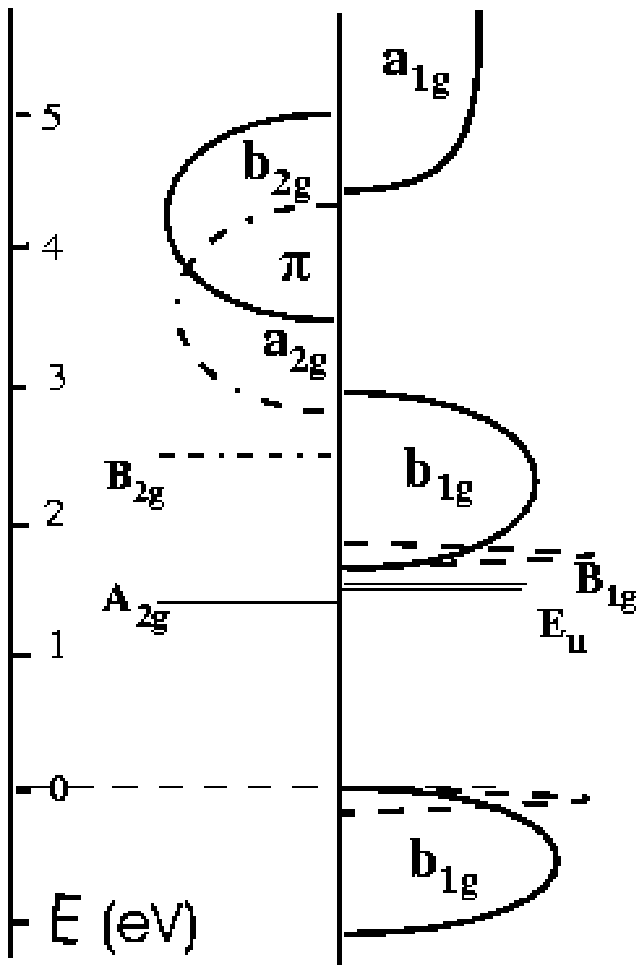,width=2.8 in,silent=}}
   \caption{
Results of a cell-perturbation calculation, reproduced from Ref. [9].
The solid lines on the right hand side represent the 
dispersion of a native hole in the insulating CuO$_2$ plane.  We assign our
5 eV feature to a transition from the $b_{1g}$ ground state to the
highest energy $a_{1g}$ excited state.
}
\end{center}
\end{figure}

\end{document}